\documentclass[%
 reprint,onecolumn,notitlepage,
 amsmath,amssymb,
 aps,
]{revtex4-1}

\usepackage{graphicx}
\usepackage[justification=justified, singlelinecheck=off, compatibility=false]{caption}
\usepackage{subcaption}
\usepackage{amsmath}
\usepackage{bm}
\usepackage{bbold}
\usepackage{amssymb}
\usepackage{xcolor}
\usepackage{times}
\usepackage{float}
\usepackage{hyperref}
\usepackage[utf8]{inputenc}
\hypersetup{
    colorlinks=true,       
    linkcolor=red,          
    citecolor=blue,        
    filecolor=magenta,      
    urlcolor=blue           
}

\newcommand{\eref}[1]{Eq.~(\ref{#1})}
\newcommand{\fref}[1]{Fig.~\ref{#1}}
\newcommand{\sref}[1]{Sec.~\ref{#1}}
\newcommand{\appref}[1]{Appendix~\ref{#1}}


\begin{document}

\title{Constraints on Neutrino Self-Interactions from IceCube Observation of NGC 1068}

\date{\today}

\author{Jeffrey M. Hyde}
\email{jhyde1@swarthmore.edu}
\affiliation{Department of Physics \& Astronomy, Swarthmore College, Swarthmore PA 19081 USA}

\begin{abstract}

The active galaxy NGC 1068 was recently identified by the IceCube neutrino observatory as the first known steady-state, extragalactic neutrino point source, associated with about 79 events over ten years. We use the IceCube data to place limits on possible neutrino self-interactions mediated by scalar particles with mass between 1 -- 10 MeV. We find that constraints on flavor-specific $\nu_{\tau}$ self-interactions with low mediator masses are comparable to constraints derived from the diffuse high-energy neutrino flux at low energies, while constraints on flavor-universal self-interactions are less restrictive than current bounds.

\vspace{1cm}
\end{abstract}

\maketitle

\setlength{\parindent}{20pt}

\section{Introduction}\label{sec:intro}

Neutrinos are an exciting component of multi-messenger astrophysics, as their weak interactions with baryonic matter can lead to inside views of distant and extreme phenomena that would otherwise be obscured from direct observation. The study of astrophysical neutrinos can also lead to insights about neutrinos themselves. In particular, neutrinos from distant sources pass through a background of Big Bang relic neutrinos and dark matter on their way to Earth. An appreciable cross section for neutrino self-interactions ($\nu$SI) or neutrino-dark matter scattering could alter the spectrum along the way, so the NGC 1068 detection allows for a test of such beyond-Standard Model interactions.

This possibility was first examined with neutrinos from Supernova 1987a \cite{Kolb:1987qy}; since then, some work has considered the ability of current and planned experiments to examine neutrino self-interactions \cite{Ng:2014pca,Shoemaker:2015qul,Creque-Sarbinowski:2020qhz,Esteban:2021tub}, motivated for example by cosmological implications \cite{Barenboim:2019tux,Blinov:2019gcj,Mazumdar:2020ibx}. Such interactions are constrained by Big Bang Nucleosynthesis (BBN), with laboratory experiments \cite{Huang:2017egl,Berryman:2022hds,Berryman:2018ogk} extending some sensitivity to higher mediator masses. IceCube results are of special interest because neutrino energies above $\sim$ TeV allow mediator masses $\gtrsim$ MeV to be probed with potentially greater sensitivity. Until recently, though, SN1987a remained the only identified point source of neutrinos. In 2017, IceCube identified a high-energy neutrino coincident with the direction and timing of the flaring blazar TXS 0506+056 \cite{IceCube:2018dnn}, and follow-up work found some evidence for neutrino emission from this source prior to the high-energy alert \cite{IceCube:2018cha}. The blazar flare event has been used to place limits on neutrino self-interactions \cite{Kelly:2018tyg}, and the diffuse high-energy neutrino flux has also been used for this purpose \cite{Ioka:2014kca,Bustamante:2020mep}.

Recently, IceCube reported a significant detection of about 79 excess neutrinos from the active galaxy NGC 1068 among data taken from 2011 and 2020, with a best-fit power law spectral index 3.2 \cite{IceCube:2022der}. For purposes of examining neutrino self-interactions, NGC 1068 has the advantage of a relatively large number of signal events and a spectrum that is assumed to be time-independent, in contrast with the flaring blazar. In contrast with SN1987a this signal is at higher energy ($\gtrsim 10^2$ GeV vs. $\sim1$ -- 10 MeV) allowing us to probe more massive mediators, it is more distant (at 14.4 Mpc versus 51.4 kpc), and it has a greater number of signal events (78 versus 22), possibly allowing more sensitivity to the coupling strength. However, the muon track events examined in point source searches such as this have weaker connection to the original neutrino energy.

In this work, we use the data released in connection with the IceCube observation of NGC 1068 \cite{IceCube:2022der,IceCube:ngc1068-data} to derive constraints on neutrino self-interactions. (We note that the TXS 0506+056 and NGC 1068 results have also been used to examine the possibility of neutrino-dark matter scattering \cite{Cline:2022qld,Ferrer:2022kei,Cline:2023tkp}.) We find that a diagonal, flavor-universal self-interaction is not constrained beyond existing bounds, while constraints on $\nu_{\tau}-\nu_{\tau}$ self-interactions are comparable to other bounds based on the diffuse neutrino flux at IceCube. \sref{sec:methods} describes our methods for modeling neutrino self-interactions and evaluating statistics, \sref{sec:results} describes our results, and in \sref{sec:conclusions} we summarize, compare with existing bounds from other sources, and describe future opportunities. \appref{sec:energy-pdf} describes in more depth how we modeled the energy pdf used in our likelihood analysis.

\section{Methods}\label{sec:methods}

\subsection{Modeling Neutrino Self-Interactions}\label{sec:flux-modeling}

In our analysis we assume that the source produces a decreasing power-law neutrino flux spectrum $\Phi \propto E_{\nu}^{-\gamma}$ over the range of observed energies, approximately $10^2$ to $10^6$ GeV, consistent with the assumption of the IceCube analysis. Neutrino self-interactions modify such a spectrum: in general, when a signal neutrino scatters, the outgoing neutrino will have diminished energy. For s-channel scattering, the cross section is largest near a resonant energy which depends on the mediator mass, and as a result a signal's power-law spectrum will have a ``dip'' near the resonant energy and an associated ``pile-up'' at lower energies. In practice, the pile-up at lower energies would be more difficult to observe, as the higher-energy part of the spectrum where they originate would have far fewer neutrinos than the lower-energy part of the spectrum where they end up. Therefore, in this work we focus on constraining dips in the spectrum from resonant s-channel scattering.

More quantitatively, we consider an interaction of the form $\mathcal{L}_{\rm int} = g \overline{\nu}\nu\phi$, where $\phi$ is a real scalar with mass $m_{\phi}$. We examine two scenarios: flavor-diagonal coupling $g$: $\mathcal{L}_{\rm int} = g \sum_i \overline{\nu}_i\nu_i\phi$, and the less constrained scenario of self-interaction only among tau neutrinos: $\mathcal{L}_{\rm int} = g \overline{\nu}_{\tau}\nu_{\tau}\phi$. Big Bang Nucleosynthesis bounds rule out mediators with masses $\lesssim 0.1$ to 1 MeV when $g \gtrsim 10^{-6}$ \cite{Ng:2014pca,Huang:2017egl,Blinov:2019gcj}. Other astrophysical and laboratory constraints are summarized in Ref. \cite{Berryman:2022hds}, and the most relevant are also plotted along with our result in \fref{fig:constraints}.

First, we gain some intuition for the result by considering resonant scattering of one neutrino flavor with no oscillations, with cross section of the Breit-Wigner form
\begin{align}\label{eq:breit-wigner}
	\sigma & = \frac{g^4}{4\pi} \frac{s}{((s-m_{\phi}^2)^2 + m_{\phi}^2\Gamma^2},
\end{align}
where $s = 2Em_{\nu}$ and the decay width is $\Gamma = g^2 m_{\phi} / (4\pi)$. Because NGC 1068 is located 14.4 Mpc from Earth (at redshift $z = 0.003$) the effect of cosmological expansion on neutrino energies and flux is negligible. The resonant energy is $E_R = m_{\phi}^2 / (2m_{\nu})$, and neutrino masses are bounded to have at least one state with mass in the range $0.05 - 0.1$ eV. For this estimate we take $m_{\nu} \approx 0.1$ eV or $10^{-10}$ GeV. A neutrino energy range of $10^2$ to $10^5$ GeV then would correspond to mediator masses in the range of about 0.1 to 4.5 MeV. However, the muon energies obtained by IceCube are of somewhat lower energy than the incident neutrino (see \sref{sec:stat-analysis} and \appref{sec:energy-pdf}), and therefore somewhat higher mediator masses can be probed as well. In any case, this estimate shows that the NGC 1068 signal should be sensitive to mediator masses above those ruled out by the BBN bound. Whether the signal is also sensitive to couplings below existing bounds is a question to be resolved by the analysis described in \sref{sec:stat-analysis}.

We now generalize to the physical case of three flavors with mixing. The IceCube result \cite{IceCube:2022der} only infers intensity at Earth of the muon-neutrino flux originating from the source; in contrast we must assume an initial flavor composition due to flavor oscillations and scattering during propagation. AGNs are expected to produce neutrinos through photomeson decay dominated by pions \cite{Stecker:1991vm}. The decays $\pi^+ \rightarrow \mu^+ + \nu_{\mu}$ and $\mu^+ \rightarrow e^+ + \nu_e + \overline{\nu}_{\mu}$ lead to an initial flavor ratio of $\nu_{\rm e} : \nu_{\mu} : \nu_{\tau} = 1:2:0$, which we use as the initial relative composition of neutrinos plus antineutrinos. We note that the most common production scenarios all lead to a roughly 1:1:1 flavor ratio after oscillations, so unless there is a scenario which modifies this our results are weakly dependent on this assumption. 

The typical distances between source, intermediate scattering (if any), and detection at Earth are far greater than the neutrino wave packets' decoherence length, so the treatment of flavor oscillations is simplified. In terms of the initial muon neutrino/antineutrino flux $\Phi_{\nu_{\mu} + \overline{\nu}_{\mu}} \equiv \Phi_{\mu}$, the fluxes of mass eigenstates $i = 1,2,3$ are
\begin{align}
	\Phi_i & = (0.5|U_{ei}|^2 + |U_{\mu i}|^2)\Phi_{\mu},
\end{align}
where $U_{\alpha i}$ are elements of the neutrino mixing matrix; we use the 2022 global best-fit parameters from NuFit \cite{Esteban:2020cvm}. We typically specify the neutrino self-coupling matrix $g_{\alpha\beta}$ in the flavor basis, but it will be computationally convenient to consider scattering in the mass basis:
\begin{align}
	g_{ij} & = \sum_{\alpha, \beta} U_{\alpha i} U_{\beta j} g_{\alpha \beta}.
\end{align}
For our first scenario, $g_{\alpha\beta} = g \ {\rm diag}(1,1,1) = g \mathbb{1}_3$, the mass-basis form $g_{ij}$ is also diagonal since the $3\times 3$ identity $\mathbb{1}_3$ commutes with $U$. For our second scenario, $g_{\alpha\beta} = g \ {\rm diag}(0,0,1)$, there will be off-diagonal contributions that couple different mass-basis states.

The generalization of the single-flavor Breit-Wigner cross section \eref{eq:breit-wigner} for signal neutrino of mass state $i$ scattering from background neutrino of mass state $j$ is \cite{Creque-Sarbinowski:2020qhz}
\begin{align}\label{eq:breit-wigner-3flavor}
	\sigma(\nu_i \nu_j \rightarrow \nu \nu) \equiv \sigma_{ij} & = \sum_{k,l} \frac{1}{S_{kl}} \frac{|g_{ij}|^2 |g_{kl}|^2}{4\pi} \frac{s_j}{((s_j-m_{\phi}^2)^2 + m_{\phi}^2\Gamma^2},
\end{align}
where $s_j = 2 E_{\nu_i} m_j$ and the decay width has been modified to $\Gamma = \sum_{i,j} |g_{ij}|^2 m_{\phi} / (4\pi)$. In \eref{eq:breit-wigner-3flavor} we have summed over the outgoing flavors $k$ and $l$, and the symmetry factor $S_{kl} = 1 + \delta_{kl}$ accounts for cases with identical particles in the final state. The number density of cosmic background neutrinos and antineutrinos is $n = 112$ cm$^{-3}$ per flavor, or 56 cm$^{-3}$ per flavor for neutrinos or antineutrinos. Based on the Planck 2018 upper bound on the sum of neutrino masses, $\sum m_{\nu} < 0.12$ eV \cite{Planck:2018vyg}, we take the heaviest neutrino masses consistent with this and oscillation-based mass-squared differences: $m_1 = 0.02$ eV, $m_2 = 0.0286$ eV, $m_3 = 0.0701$ eV.

The neutrino optical depth for mass state $i$ traveling distance $D$ from source to detector is
\begin{align}
	\tau_i & = \frac{D}{\lambda_i(E)} \ = \ D \sum_j \sigma_{ij}(E) \ n_j,
\end{align}
and the associated mass-basis neutrino flux is attenuated from its original value $\Phi^0_i(E)$ to become
\begin{align}\label{eq:mass-flux-earth}
	\Phi_i(E) & = \exp(-\tau_i(E)) \Phi^0_i(E).
\end{align}
Finally, the transformation
\begin{align}\label{eq:muon-flux-earth}
	\Phi_{\mu}(E) & = \sum_i |U_{\mu i}|^2 \Phi_i(E)
\end{align}
gives the flux spectrum of muon neutrinos incident at Earth.

\subsection{Statistical Analysis}\label{sec:stat-analysis}

The IceCube observation of NGC 1068 reported in \cite{IceCube:2022der} is obtained from a dataset of 19,452 neutrino events within 15 degrees in right ascension (RA) and declination (DEC) of NGC 1068 (RA = $40.667^{\circ}$, DEC = $-0.0067^{\circ}$), representing 3186 days of livetime from 2011 to 2020. These muon track events have good angular resolution and larger statistics, but at the cost of less knowledge of initial neutrino energy; overall, the dataset consists of atmospheric neutrino background plus possible signal. Assuming a power law signal $\Phi = \Phi_0 (E/E_0)^{-\gamma}$, a likelihood ratio test led to the quoted result of $\Phi_0 = (5.0 \pm 1.5) \times 10^{-11}$ TeV$^{-1}$ cm$^{-2}$ s$^{-1}$ at $E_0 = 1$ TeV, $\gamma = 3.2 \pm 0.2$, or $79^{+22}_{-20}$ signal events.

To examine the possibility of neutrino self-interactions, we consider the data to be comprised of atmospheric neutrinos along with a power-law spectrum from NGC 1068 which is modified as described in \sref{sec:flux-modeling}. For data $x_i$ (where $x_i = \{ \rm Energy, \ Declination, \ Ang. \ Unc.\} \equiv \{ E_i, \ \delta_i, \ \sigma_i\}$) and model parameters $\theta_i$ (where $\theta_i = \{ \gamma, n_s, g, m_{\phi} \}$) we take the likelihood function to be \cite{Braun:2008bg}
\begin{align}\label{eq:likelihood}
	\mathcal{L}( \{x_i \} | \{ \theta_i\} ) & = \prod_{i=1}^{N'} \left[ \frac{n_s}{N} f_{\rm signal}(x_i | \theta_i ) + \left( 1 - \frac{n_s}{N} \right) f_{\rm background}(x_i) \right],
\end{align}
where the atmospheric neutrino spectrum is described by $f_{\rm background}(E)$ and is identical to that used in \cite{IceCube:2022der}. The number $N = 665,293$ represents the number of events in the all-sky sample, and $N' = 19,452$ is the number in the sample within 15 degrees of NGC 1068. In principle the product is over all $N$ events, but we follow the IceCube analysis in assuming there are no signal events from NGC 1068 at greater than 15 degrees. The signal probability density function (pdf) $f_{\rm signal}$ is the product
\begin{align}\label{eq:signal-pdf}
	f_{\rm signal}(x_i) & \approx \frac{1}{2\pi\sin(\hat{\psi})} f_{\rm energy}(E | \sin\delta, \gamma, \mu_{\rm ns}, m_{\phi}, g) f_{\rm spatial}(\psi_i | E, \sigma, \sin{\delta}, \gamma),
\end{align}
where $\psi$ is the angular separation between a given neutrino event and the source coordinates. Here we carry over the approximation made in the IceCube analysis that an angular error component may be dropped from the signal and background pdfs due to weak dependence on $\gamma$. Furthermore, since we only consider NGC 1068, we drop explicit reference to declination in each function. When $m_{\phi}$ and $g$ vanish, $f_{\rm energy}$ reduces to the function used in the IceCube analysis. Our modeling of the modified energy pdf in the presence of $\nu$SI is discussed in detail in \appref{sec:energy-pdf}.

Neutrino self-interactions can lead to ``echo'' signals appearing to come from a displaced origin, but the highly forward scattering of neutrinos at these energies means that such echoes would occur at negligible angles from the original source, with typical scattering angles $\lesssim 10^{-7}$ for 100 TeV neutrinos \cite{Murase:2019xqi}. Therefore, we take the spatial pdf to be independent of neutrino self-interactions. Because the atmospheric neutrinos that make up the background are not propagating an appreciable distance (or optical depth) between production and detection, we also take the background pdf to be independent of neutrino self-interactions. Therefore, in both cases we adopt the spatial pdf for the signal and background pdf from IceCube.

We take as test hypothesis (hereafter labeled $H_1$) the presence of neutrino self-interactions with $g\neq 0$, and as null hypothesis ($H_0$) the case $g = 0$ and combination of background and signal with parameters given by the best fit IceCube result, namely $\gamma^{\ast} = 3.26$, $n_s^{\ast} = 79$. As test statistic we define the log-likelihood ratio
\begin{align}\label{eq:llh-def}
	\lambda & \equiv 2\log\left( \frac{\mathcal{L}_{H1}}{\mathcal{L}_{H0}} \right).
\end{align}
Large values of $\lambda$ favor the test hypothesis $H_1$, and in the large-sample limit this test statistic should follow a chi-square distribution \cite{Cowan:1998ji}, which we will use in \sref{sec:results} to set confidence regions.

\section{Results}\label{sec:results}

\begin{figure}[t]
	\centering
	\begin{subfigure}[b]{0.45\textwidth}
		\centering
		\includegraphics[width=\textwidth]{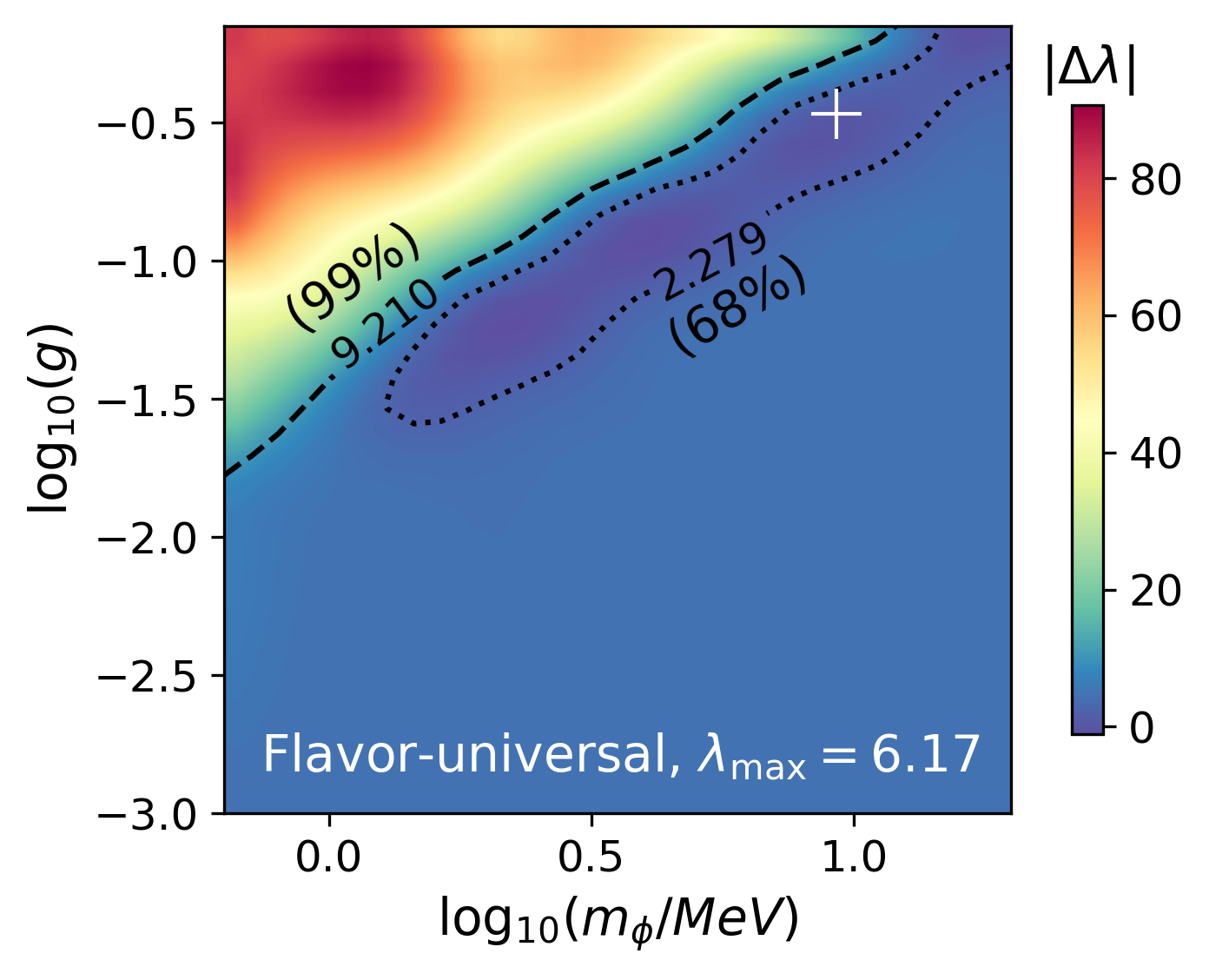}
		\caption{\label{fig:m-g-scan-1}}
	\end{subfigure} \, \, \, \, \, \, %
	\begin{subfigure}[b]{0.45\textwidth}
		\centering
		\includegraphics[width=\textwidth]{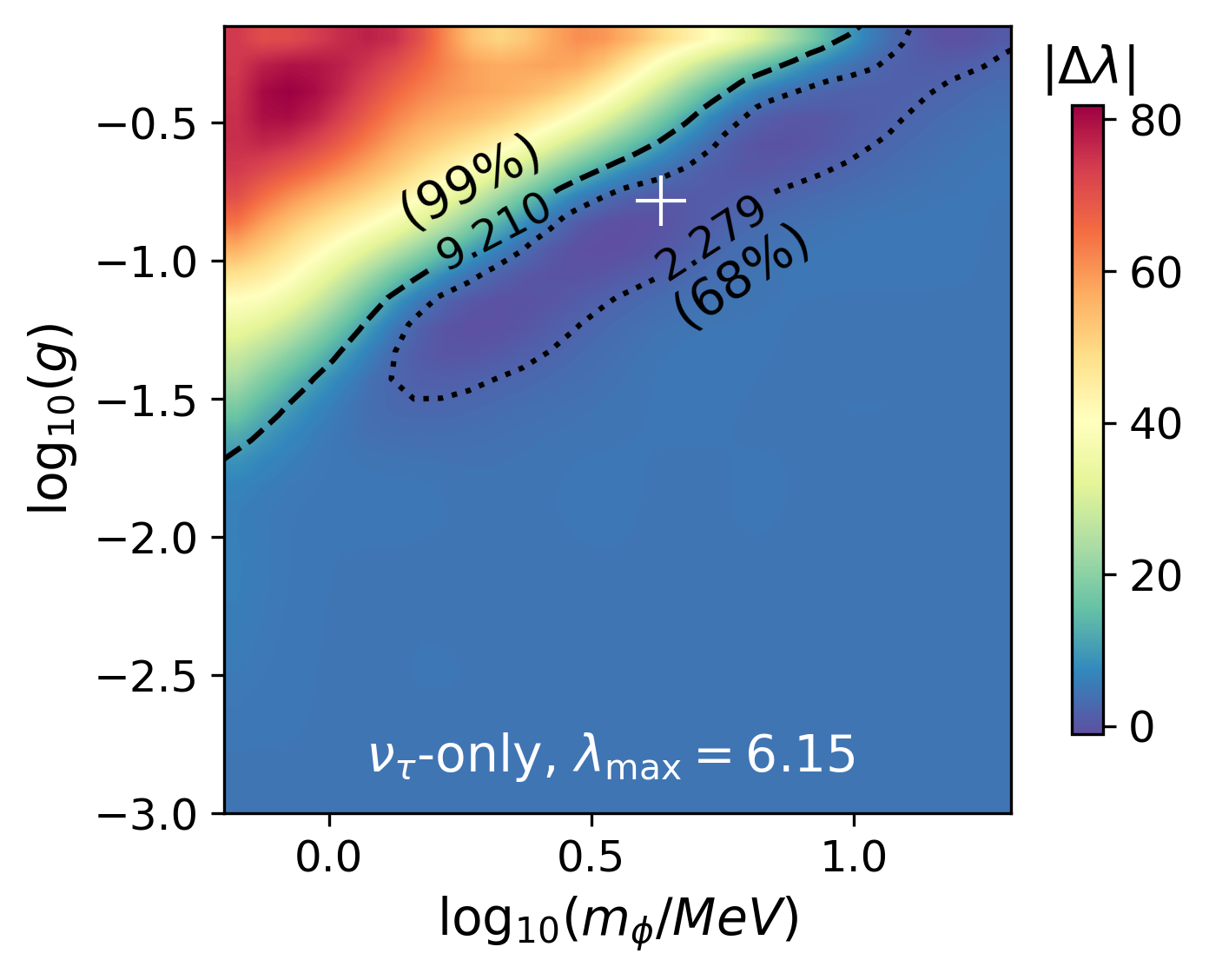}
		\caption{\label{fig:m-g-scan-2}}
	\end{subfigure}
\caption{\label{fig:m-g-scan} Results of likelihood analysis in the parameter space of coupling ($g$) and mediator mass ($m_{\phi}$), for (\fref{fig:m-g-scan-1}) flavor-universal neutrino self-interaction and (\fref{fig:m-g-scan-2}) only $\nu_{\tau}-\nu_{\tau}$ self-interaction. Both show the maximum value of the test statistic $\lambda$ found at a given $g$, $m_{\phi}$ across a scan of spectral indices ($\gamma$) and number of signal events ($n_s$) as described in the text. The ``+'' sign marks the location with highest value of the test statistic $\lambda = 2\log(L_1/L_0)$, while the curves show deviations of $\lambda$ representing different confidence levels, as labeled on the figure and found as described in the text.} 
\end{figure}

We evaluated the test statistic $\lambda$ defined in \eref{eq:llh-def} on a grid of 19,500 points in parameter space, comprised of $N_g = 10$ values of $\log_{10}(g)$ (from $-3$ to $-0.15$), $N_m = 10$ values of $\log_{10}(m_{\phi}/{\rm MeV})$ (from $-0.2$ to $+1.3$), $N_{\gamma} = 13$ values of $\gamma$ (from $2.8$ to $3.6$), and $N_n = 15$ values of $n_s$ (from 35 to 125). The values of $m$ and $g$ were chosen to focus on the parameter space not covered by lab and BBN bounds, and to include the point $g = 0.10$, $m_{\phi} = 14$ MeV, where the likelihood was maximized (though with low significance) in \cite{Bustamante:2020mep}.

We then use a spline interpolation of the test statistic, with results shown in \fref{fig:m-g-scan} for both scenarios. In \fref{fig:m-g-scan} we plot the magnitude of the difference $|\Delta \lambda|$ between the test statistic at a given point and the overall test-statistic maximum, whose value is listed. The significance of the test result, and confidence regions, are evaluated using a 2-parameter chi-square distribution, representing the difference in dimensionality between the parameter space of null and test hypotheses. We treat the spectral index, $\gamma$, and the number of signal events, $n_s$, as nuisance parameters. For a given pair $g$, $m_{\phi}$ we take the values of $\gamma$ and $n_s$ which together maximize the test statistic.

For the flavor-universal case, the maximum value of $\lambda$ is 6.17 (p-value 0.0457), occurring at $g = 0.341$, $m_{\phi} = 9.26$ MeV, and indicated by a $``+''$ in \fref{fig:m-g-scan-1}. For the $\nu_{\tau}-\nu_{\tau}$ case, the maximum value of $\lambda$ is 6.15 (p-value 0.0462), occurring at $g = 0.165$, $m_{\phi} = 4.30$ MeV. In each case, there is insufficient evidence to reject the null hypothesis. We can see from the plot that $\lambda$ is essentially constant for a large region but decreases significantly as $g$ increases and $m_{\phi}$ decreases; we use the 99\% confidence region to set bounds in \sref{sec:conclusions}. The similarity between the results for each case reflects energy smearing and flavor mixing built into the energy pdf.

\begin{figure}[t]
	\centering
	\begin{subfigure}[b]{0.45\textwidth}
		\centering
		\includegraphics[width=\textwidth]{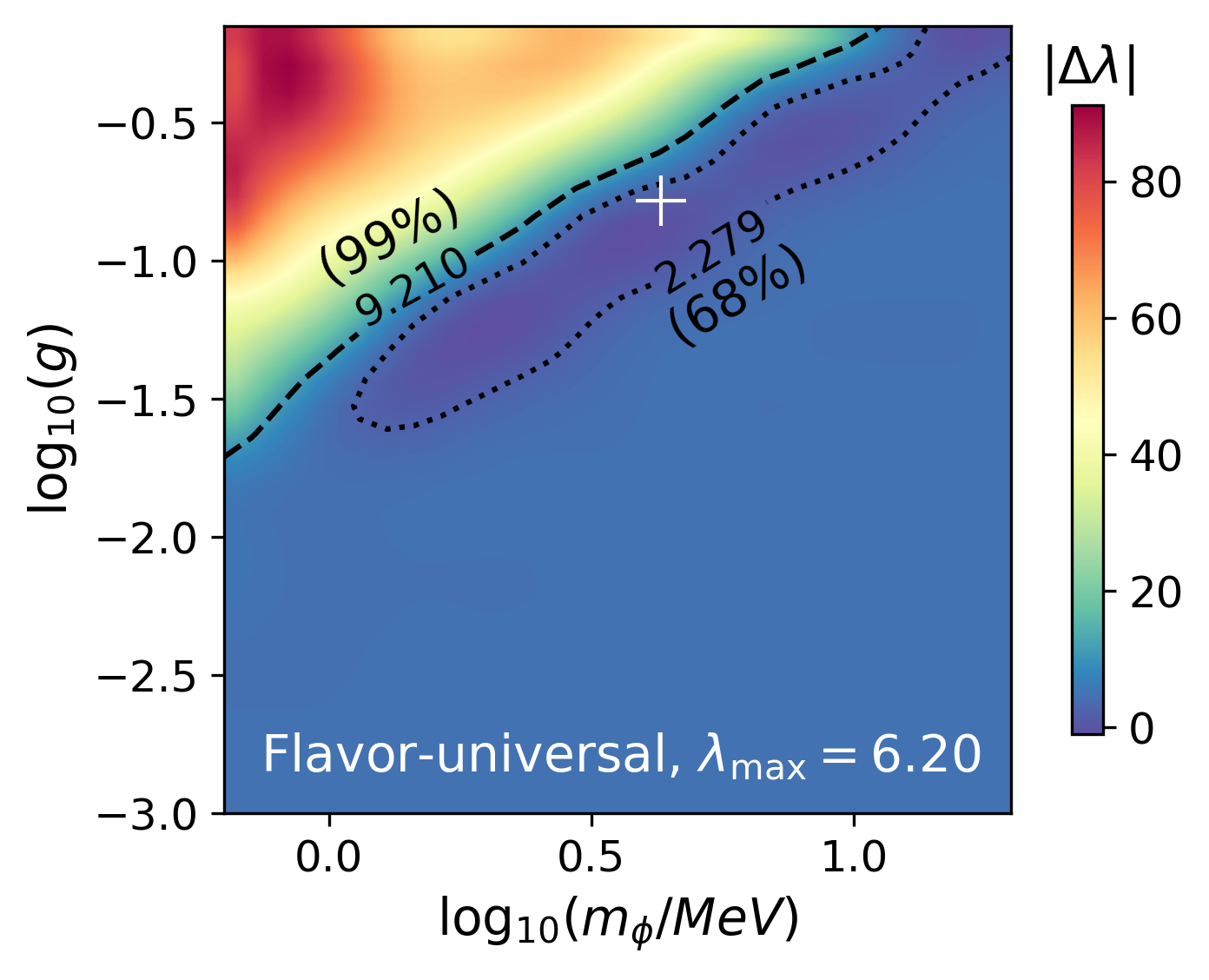}
		\caption{\label{fig:m-g-scan-1-lowermass}}
	\end{subfigure} \, \, \, \, \, \, %
	\begin{subfigure}[b]{0.45\textwidth}
		\centering
		\includegraphics[width=\textwidth]{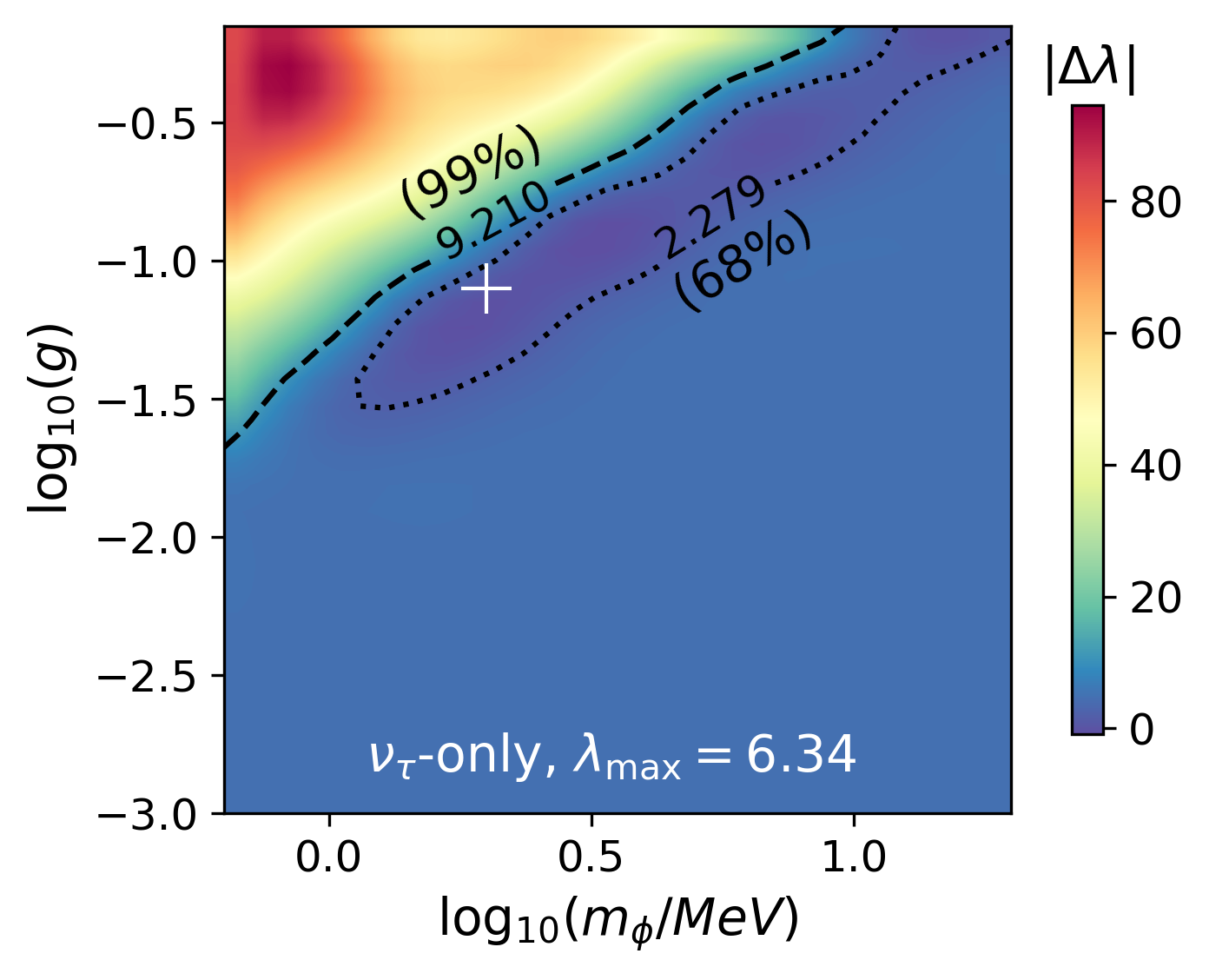}
		\caption{\label{fig:m-g-scan-2-lowermass}}
	\end{subfigure}
\caption{\label{fig:m-g-scan-lowermass} Same as \fref{fig:m-g-scan}, but for a different choice of neutrino mass values $m_i$, as described in the text. The uncertainty in incident neutrino energy encapsulated in the energy pdfs (see \appref{sec:energy-pdf}) renders this difference in neutrino mass scale insignificant, and the bounds are qualitatively unchanged.}
\end{figure}

We also test the effect of varying the neutrino mass within the bounds discussed in \sref{sec:flux-modeling}. The above results take the the heaviest allowed mass scale, with $m_1$'s range limited to $[0, 0.02]$ eV in the normal ordering. Here we subtract 0.01 eV from all $m_i$, half of the available range in $m_i$'s given the bounds on $m_1$, and repeat the above analysis, with results plotted in \fref{fig:m-g-scan-lowermass}. For Case 1, $\lambda_{\rm max} = 6.20$ at $m_{\phi} = 4.30$ MeV, $g = 0.16$. For Case 2, $\lambda_{\rm max} = 6.34$ at $m_{\phi} = 2.01$ MeV, $g = 0.08$. While the significance of some regions of parameter space have changed moderately, the boundary of the 99\% confidence region is nearly unchanged. Qualitatively, we attribute this to the energy smearing and flavor mixing mentioned above, which render the details of these resonance structures indistinguishable in the energy pdf. This change in the neutrino mass would also affect the resonant energies, but evidently not enough to significantly affect the outcome.

\section{Discussion and Conclusions}\label{sec:conclusions}

In this paper we have used the data release accompanying the recent IceCube detection of neutrinos from NGC 1068 \cite{IceCube:2022der,IceCube:ngc1068-data} to place limits on neutrino self-interactions. The results presented in \fref{fig:m-g-scan} are similar for the flavor-diagonal versus $\nu_{\tau}$-specific cases, with flavor-diagonal providing a slightly stronger bound. While different self-interaction matrices $g_{\alpha\beta}$ lead to different resonance structure in the incident flux, the significant amount of energy smearing in the muon track energy pdf renders this information inaccessible, so the pdfs -- and therefore results of the statistical analysis -- are very similar.

\begin{figure}[t]
	\centering
	\begin{subfigure}[b]{0.45\textwidth}
		\centering
		\includegraphics[width=\textwidth]{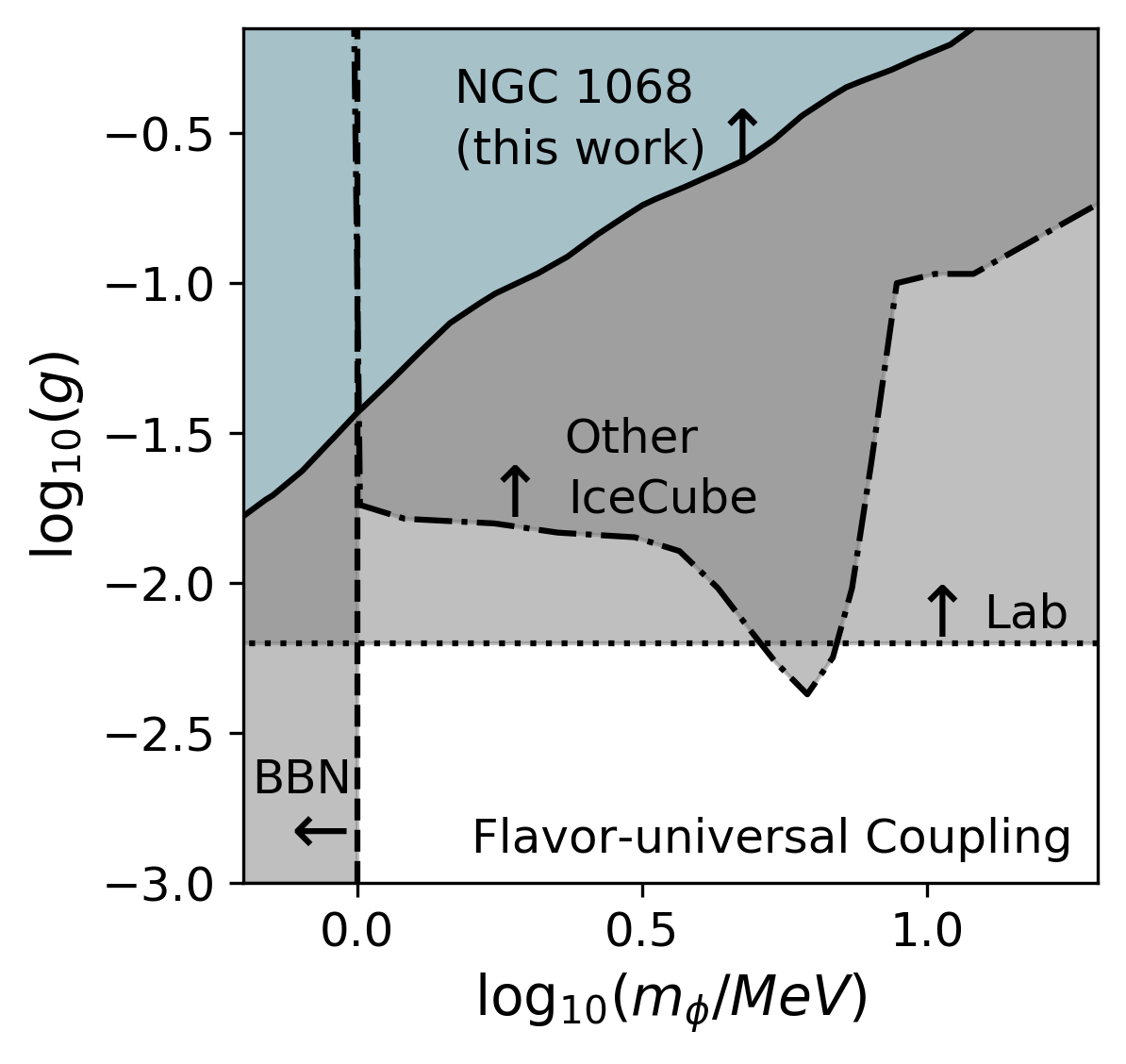}
		\caption{\label{fig:constraint-case1}}
	\end{subfigure} \, \, \, \, \, \, %
	\begin{subfigure}[b]{0.45\textwidth}
		\centering
		\includegraphics[width=\textwidth]{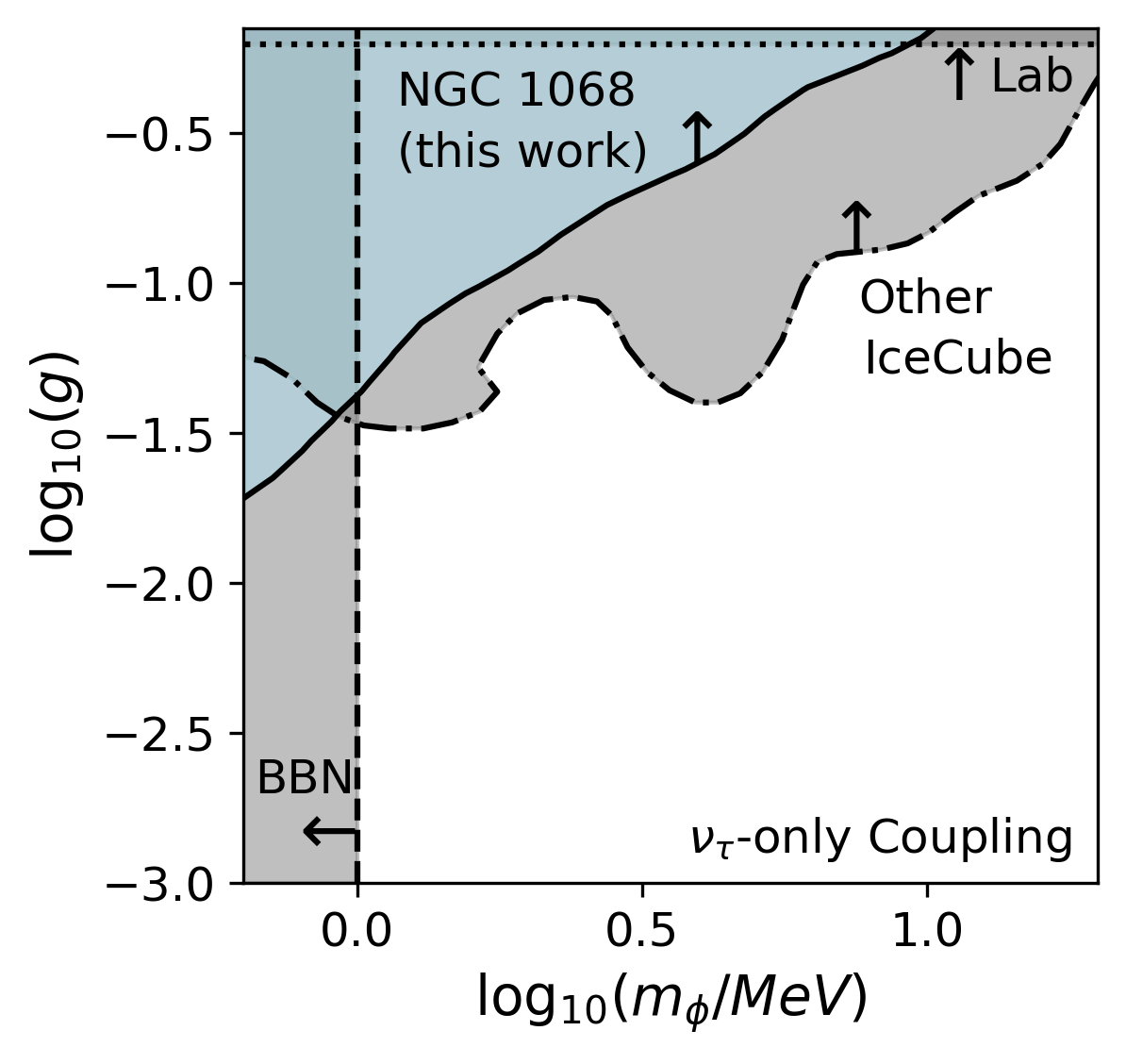}
		\caption{\label{fig:constraint-case2}}
	\end{subfigure}
\caption{\label{fig:constraints} Comparison of our constraint curve based on the boundary of the 99\% confidence region with existing bounds on neutrino self-interactions, for (a) flavor-universal self-interactions, and (b) $\nu_{\tau}-\nu_{\tau}$ self-interactions. Constraints are taken from the summary in Ref. \cite{Berryman:2022hds}, as well as \cite{Esteban:2021tub} for IceCube HESE constraints on $\nu_{\tau}$-specific self-interactions.}
\end{figure}

The similar flavor-universal and $\nu_{\tau}$-specific results have different interpretations in relation to existing bounds. Flavor-universal self-interactions are both more accessible experimentally and more often examined in relation to results from IceCube and other experiments. For example, lab bounds on flavor-universal self-interactions include kaon and pion decay constraints on electron-flavor self-interactions, while bounds on the $\nu_{\tau}$-only case from Higgs and Z decays are less restrictive \cite{Berryman:2022hds,Berryman:2018ogk}. As a result, while our flavor-universal results are not as strong as existing bounds, our $\nu_{\tau}$-specific results are much stronger than lab-based, and comparable (especially at lower energies) to those derived from the IceCube diffuse flux, as seen in \fref{fig:constraints}. The constraints plotted in \fref{fig:constraints} all have differences in assumptions and methodology, and a joint analysis of the IceCube data could be valuable.

In \sref{sec:intro}, we noted that the high-energy starting event (HESE) sample has fewer events and more uncertainty in neutrino direction, but less uncertainty in the incident neutrino energy. On balance, the comparison \fref{fig:constraint-case1} demonstrates that the present point-source data from NGC 1068 does not outweigh the sensitivity of the HESE data. However, the analysis leading to the discovery of NGC 1068 as a source of high-energy neutrinos also tested a number of other active galaxies and blazars; while none yet reach the same discovery threshold, a few had suggestive excesses of $\sim 3.5 \sigma$. With more years of data taking and greater sensitivity of future upgrades, it is possible that a larger point source dataset could lead to significantly tighter constraints.

As described in \appref{sec:energy-pdf}, modeling of IceCube's energy pdf is necessary for this analysis, but challenging in general. We are fortunate here that BBN constraints generally restrict our new physics effects to the easier-to-model high energy end of the spectrum, but future analyses for which new physics could affect the low-energy end of the spectrum would have to improve upon this. Furthermore, the assumption shared by our analysis and IceCube of a strict power-law spectrum for neutrinos from AGN emission is unlikely to be a precise representation of the physics across 4 decades of energy. In fact, for larger mediator masses the energy pdfs in \sref{sec:energy-pdf} look qualitatively similar to what one might expect if the true spectrum is a broken power law, and the moderate preference for $\nu$SI along a diagonal curve in the $m_{\phi}-g$ plane (\fref{fig:m-g-scan-2}) could be the result of such a degeneracy. With more data from neutrino and other observatories, future improvements to modeling of AGN neutrino spectra \cite{Murase:2022feu,Creque-Sarbinowski:2021nil} could in turn lead to different constraints on neutrino self-interactions.

While this work was in preparation, another paper appeared using different methodology to address the question of NGC 1068 bounds on neutrino self-interactions \cite{Doring:2023vmk}. In contrast with our unbinned likelihood analysis of the data, their work compares calculations of binned event rates in the presence or absence $\nu$SI (not restricted to resonant s-channel interactions), based on IceCube's effective area but not the NGC 1068 data. Within the parameter space region where our analyses overlap, Ref. \cite{Doring:2023vmk} finds a constraint curve for flavor-universal self-interactions that is qualitatively similar in appearance to \fref{fig:m-g-scan-1} but with a weaker bound.

\acknowledgments

I am very grateful to William Luszczak, Peter Denton and Tristan Smith for useful conversations. This work used data and pdfs, as well as code for interfacing with these, from the IceCube NGC 1068 data release \cite{IceCube:ngc1068-data}.

\begin{appendix}

\section{Energy PDF}\label{sec:energy-pdf}

Here we provide further details regarding our modeling of the energy probability density function (pdf) -- $f_{\rm energy}$ referenced in \eref{eq:signal-pdf} -- used in our statistical analysis. For convenience, we work in energy parameters $\epsilon \equiv \log_{10}(E/{\rm GeV})$, where the incident neutrino has $\epsilon_{\nu} = \log(E_{\nu})$ and the secondary muon has true energy $\epsilon_{\mu} = \log(E_{\mu})$ and reconstructed (observed) energy $\hat{\epsilon}_{\mu} = \log(\hat{E}_{\mu})$.

IceCube provides effective areas as well as the overall energy pdf in the absence of neutrino self-interactions, $f(\hat{\epsilon}_{\mu}|\gamma)$, which they have obtained via Monte Carlo simulation. These simulations include many effects which are impractical to include in a phenomenological analysis such as ours, for instance, detailed modeling of photodetector response and the propagation of Cherenkov light in Antarctic ice. However, by incorporating the dominant physical processes of muon energy loss and energy reconstruction uncertainty, the energy pdf can be recreated sufficiently well, as will be shown below.

Formally, the energy pdf can be expressed as (see e.g. \cite{Braun:2008bg,Braun:thesis})
\begin{align}\label{eq:epdf-formal}
	f(\hat{\epsilon}_{\mu}|\gamma,g,m_{\phi}) & = N^{-1} \int d\epsilon_{\nu} \ P_{\rm prop.}(\hat{\epsilon}_{\mu} | \epsilon_{\nu}) \ P_{\rm int.}(\epsilon_{\nu}|\gamma,g,m_{\phi}),
\end{align}
where $P_{\rm int.}(\epsilon_{\nu} | \gamma, g, m_{\phi})$ is the relative probability of a neutrino having energy $\epsilon_{\nu}$ (among those which interact and also produce muons that reach the detector), $P_{\rm prop.}(\hat{\epsilon}_{\mu} | \epsilon_{\nu})$ is the probability of obtaining reconstructed muon energy $\hat{\epsilon}_{\mu}$ given neutrino of energy $\epsilon_{\nu}$, and $N$ is a normalization factor. For this analysis we will normalize the final pdf on the interval $\hat{\epsilon}_{\mu} \in [2,6]$; therefore in the following discussion we drop overall multiplicative constants that would ultimately be absorbed into the definition of $N$. We also note that in principle \eref{eq:epdf-formal} should include directional dependence; since we are only considering one source candidate we suppress this notation for ease of reading.

We can write $P_{\rm int.}$ as
 \begin{align}\label{eq:p-int}
	P_{\rm int.}(\epsilon_{\nu}|\gamma,g,m_{\phi}) & \propto 10^{\epsilon_{\nu}} \Phi(\epsilon_{\nu} | \gamma, g, m_{\phi}) A_{\rm eff.}(\epsilon_{\nu}),
\end{align}
where $A_{\rm eff.}$ is the detector effective area for muon neutrino charged-current interactions along the incident direction. We include the Jacobian factor $|dE_{\nu}/d\epsilon_{\nu}| = \ln(10) 10^{\epsilon_{\nu}} \rightarrow 10^{\epsilon_{\nu}}$ to ensure that we obtain a pdf in $\epsilon_{\nu}$ rather than $E_{\nu}$.

We now turn to $P_{\rm prop.}(\hat{\epsilon}_{\mu} | \epsilon_{\nu})$, the probability that a neutrino of energy $\epsilon_{\nu}$ (among those which do interact and produce a muon that passes through the instrumented volume) leads to a muon whose energy is reconstructed to be $\hat{\epsilon}_{\mu}$. We use tabulated IceCube values for this function (see Fig. 4 of \cite{IceCube:2021xar} and tabulated data in the associated data release); these represent the result of extensive Monte Carlo simulation and are provided in half-decade energy bins. We use a spline interpolation, taking the given values at the bin center, to evaluate $P_{\rm prop.}$ in \eref{eq:epdf-formal}.

\begin{figure}[t]
	\centering
	\begin{subfigure}[b]{0.47\textwidth}
		\centering
		\includegraphics[width=\textwidth]{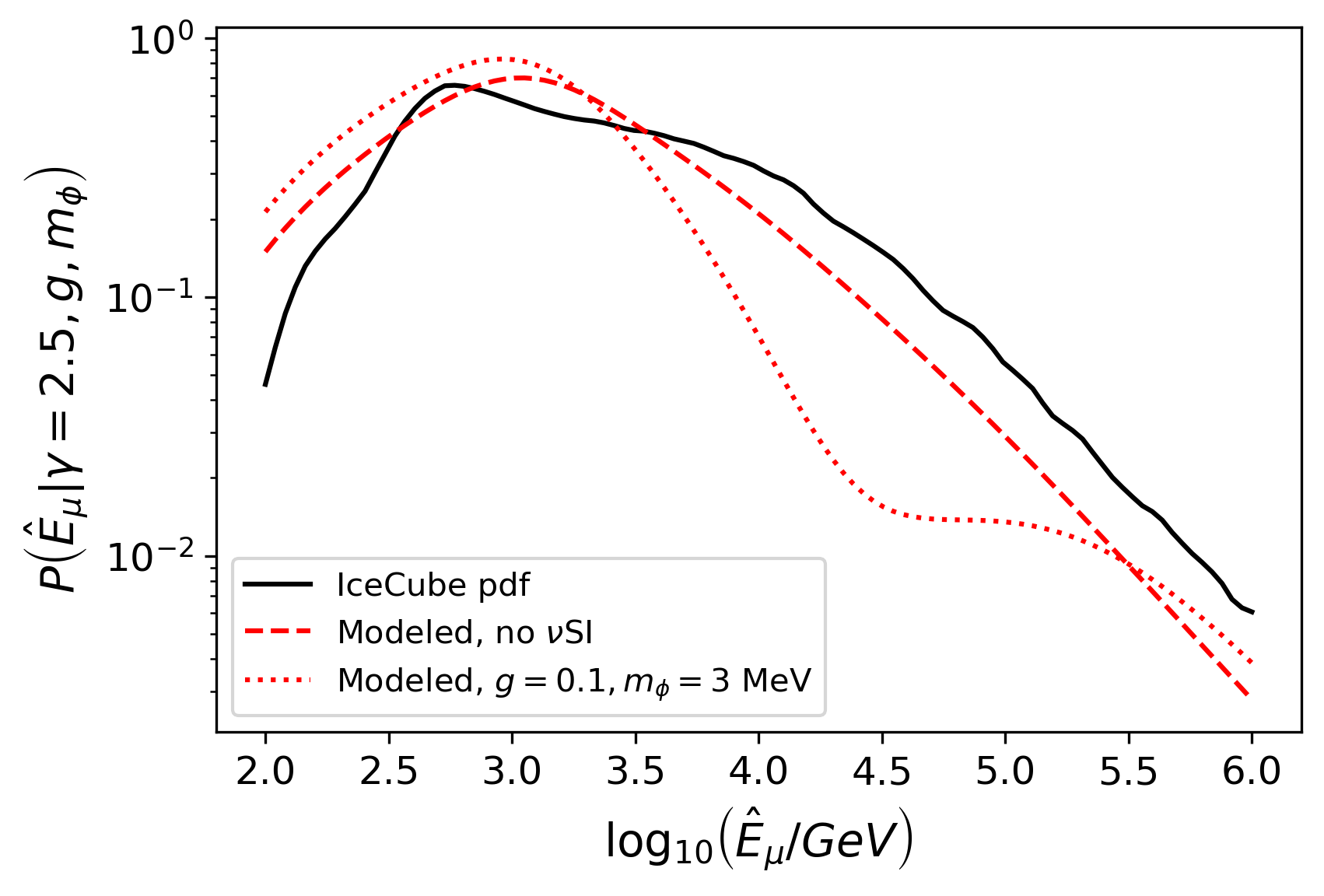}
		\caption{\label{fig:epdf-gamma2p5}}
	\end{subfigure} \, \, \, \, \, \, %
	\begin{subfigure}[b]{0.47\textwidth}
		\centering
		\includegraphics[width=\textwidth]{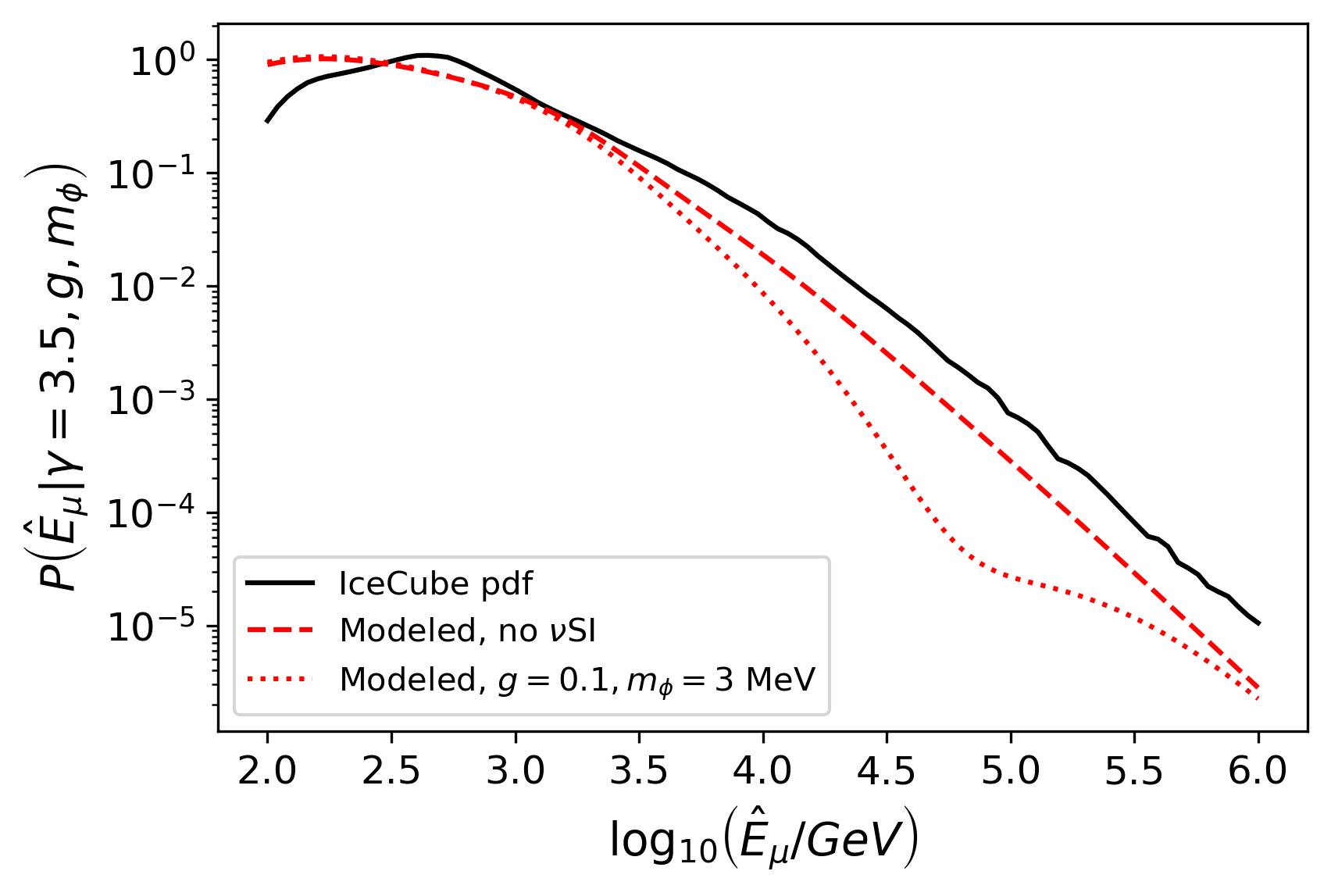}
		\caption{\label{fig:epdf-gamma3p5}}
	\end{subfigure}
\caption{\label{fig:no-nusi-epdfs} Comparison of our computed energy pdfs, in the case of no neutrino self-interactions, with those provided by IceCube, for spectral index $\gamma = 2.5$ in \fref{fig:epdf-gamma2p5} and $\gamma = 3.5$ in \fref{fig:epdf-gamma3p5}. The close match of the slope at high energy gives us confidence that our modeling of the $\nu SI$ energy pdfs reliably account for muon energy loss and reconstruction uncertainty at the detector. Discrepancies at lower energies are not significant for our analysis, as BBN bounds on mediator masses prevent us from having to consider resonances at the low-energy end of the spectrum. The dotted lines show the effect of neutrino self-interactions on the modeled pdfs, in this case using $g = 0.1$, $m_{\phi} = 3$ MeV.}
\end{figure}

\fref{fig:no-nusi-epdfs} shows the results of our modeling, in comparison to the IceCube pdfs, for two choices of spectral index ($\gamma$ = 2.5 and 3.5). We also show an example of the effect of $\nu$SI, for the choice $g = 0.1$, $m_{\phi} = 3$ MeV. Note that the distinct resonant dips in the flux due to different neutrino mass states are smoothed together by the uncertainty in incident neutrino energy. While the modeling is less precise at low energies, BBN bounds on mediator masses restrict effects of $\nu$SI to higher energies, where the slope of the pdf matches very well. We note that the normalization factor is influenced by the modeling of low-energy physics, but does not affect the final version of the pdf that we use, as discussed below.

\begin{figure}[t]
	\centering
	\begin{subfigure}[b]{0.47\textwidth}
		\centering
		\includegraphics[width=\textwidth]{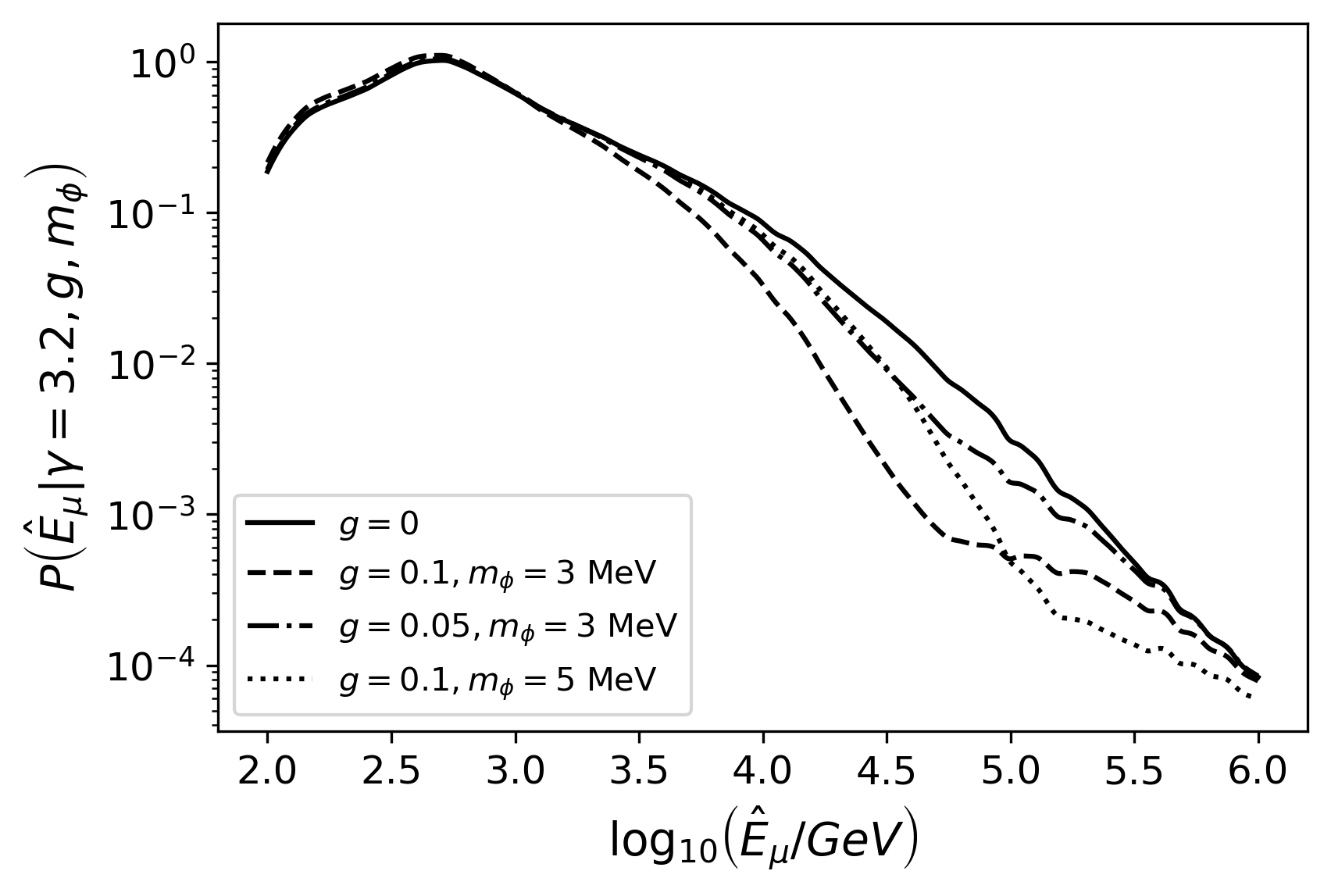}
	\end{subfigure}
\caption{\label{fig:full-epdf-examples} Example of the final version of energy pdfs, evaluated for the best-fit spectral index $\gamma = 3.2$ of the original IceCube analysis, and for several examples of $\nu$SI parameters. These have been computed as described in the text, where at each energy the ratio of the modeled pdf with $\nu$SI to that without $\nu$SI is applied to the original IceCube pdf.} 
\end{figure}

Finally, we use these modeled results to obtain the energy pdf used in our analysis in the following way. For a given choice of parameters $g$, $m_{\phi}$, at each energy we find the ratio $R(\hat{\epsilon}_{\mu} | \gamma, g, m_{\phi}) \equiv P(\hat{\epsilon}_{\mu} | \gamma, g, m_{\phi}) / P(\hat{\epsilon}_{\mu} | \gamma, 0, 0)$, then multiply this ratio by the original IceCube pdf: $f_{\rm energy}(\hat{\epsilon}_{\mu} | \gamma, g, m_{\phi}) = R(\hat{\epsilon}_{\mu} | \gamma, g, m_{\phi}) f_{\rm I.C.}(\gamma)$. Examples of the resulting energy pdfs are shown in \fref{fig:full-epdf-examples}. While our modeling ensures that we correctly account for the effect of detector physics on the resonant dips in the spectrum, this ratio ensures that in the limit $g\rightarrow 0$ we regain the exact IceCube pdfs and therefore the same best-fit parameters in the absence of $\nu$SI. This ratio method also ensures that the normalization factor for the modeled pdf, which is influenced by the less-precise low-energy modeling, does not ultimately affect the outcome of the analysis.

Overall, the loss of energy information due to propagation and reconstruction affects the strength of conclusions one can draw about self-interactions. Our modeling of the energy pdf therefore builds into the analysis an accounting of this effect. Furthermore, since not all events contributed equally to the signal in IceCube's original result it follows that the sensitivity to neutrino self-interactions is uneven across the $10^2$ to $10^{5.2}$ GeV energy range encompassing the observed muon energies, another effect that our methodology accounts for.

\end{appendix}


\begin{thebibliography}{99}


\bibitem{Kolb:1987qy}
E.~W.~Kolb and M.~S.~Turner,
``Supernova SN 1987a and the Secret Interactions of Neutrinos,''
Phys. Rev. D \textbf{36}, 2895 (1987)
doi:10.1103/PhysRevD.36.2895

\bibitem{Shoemaker:2015qul}
I.~M.~Shoemaker and K.~Murase,
``Probing BSM Neutrino Physics with Flavor and Spectral Distortions: Prospects for Future High-Energy Neutrino Telescopes,''
Phys. Rev. D \textbf{93}, no.8, 085004 (2016)
doi:10.1103/PhysRevD.93.085004
[arXiv:1512.07228 [astro-ph.HE]].

\bibitem{Creque-Sarbinowski:2020qhz}
C.~Creque-Sarbinowski, J.~Hyde and M.~Kamionkowski,
``Resonant neutrino self-interactions,''
Phys. Rev. D \textbf{103}, no.2, 023527 (2021)
doi:10.1103/PhysRevD.103.023527
[arXiv:2005.05332 [hep-ph]].

\bibitem{Esteban:2021tub}
I.~Esteban, S.~Pandey, V.~Brdar and J.~F.~Beacom,
``Probing Secret Interactions of Astrophysical Neutrinos in the High-Statistics Era,''
[arXiv:2107.13568 [hep-ph]].

\bibitem{Ng:2014pca}
K.~C.~Y.~Ng and J.~F.~Beacom,
``Cosmic neutrino cascades from secret neutrino interactions,''
Phys. Rev. D \textbf{90}, no.6, 065035 (2014)
[erratum: Phys. Rev. D \textbf{90}, no.8, 089904 (2014)]
doi:10.1103/PhysRevD.90.065035
[arXiv:1404.2288 [astro-ph.HE]].

\bibitem{Barenboim:2019tux}
G.~Barenboim, P.~B.~Denton and I.~M.~Oldengott,
``Constraints on inflation with an extended neutrino sector,''
Phys. Rev. D \textbf{99}, no.8, 083515 (2019)
doi:10.1103/PhysRevD.99.083515
[arXiv:1903.02036 [astro-ph.CO]].

\bibitem{Blinov:2019gcj}
N.~Blinov, K.~J.~Kelly, G.~Z.~Krnjaic and S.~D.~McDermott,
``Constraining the Self-Interacting Neutrino Interpretation of the Hubble Tension,''
Phys. Rev. Lett. \textbf{123}, no.19, 191102 (2019)
doi:10.1103/PhysRevLett.123.191102
[arXiv:1905.02727 [astro-ph.CO]].

\bibitem{Mazumdar:2020ibx}
A.~Mazumdar, S.~Mohanty and P.~Parashari,
``Flavour specific neutrino self-interaction: H $_{0}$ tension and IceCube,''
JCAP \textbf{10}, 011 (2022)
doi:10.1088/1475-7516/2022/10/011
[arXiv:2011.13685 [hep-ph]].

\bibitem{Huang:2017egl}
G.~y.~Huang, T.~Ohlsson and S.~Zhou,
``Observational Constraints on Secret Neutrino Interactions from Big Bang Nucleosynthesis,''
Phys. Rev. D \textbf{97}, no.7, 075009 (2018)
doi:10.1103/PhysRevD.97.075009
[arXiv:1712.04792 [hep-ph]].

\bibitem{Berryman:2022hds}
J.~M.~Berryman, N.~Blinov, V.~Brdar, T.~Brinckmann, M.~Bustamante, F.~Y.~Cyr-Racine, A.~Das, A.~de Gouv\^ea, P.~B.~Denton and P.~S.~B.~Dev, \textit{et al.}
``Neutrino Self-Interactions: A White Paper,''
[arXiv:2203.01955 [hep-ph]].

\bibitem{Berryman:2018ogk}
J.~M.~Berryman, A.~De Gouv\^ea, K.~J.~Kelly and Y.~Zhang,
``Lepton-Number-Charged Scalars and Neutrino Beamstrahlung,''
Phys. Rev. D \textbf{97}, no.7, 075030 (2018)
doi:10.1103/PhysRevD.97.075030
[arXiv:1802.00009 [hep-ph]].

\bibitem{IceCube:2018dnn}
M.~G.~Aartsen \textit{et al.} [IceCube, Fermi-LAT, MAGIC, AGILE, ASAS-SN, HAWC, H.E.S.S., INTEGRAL, Kanata, Kiso, Kapteyn, Liverpool Telescope, Subaru, Swift NuSTAR, VERITAS and VLA/17B-403],
``Multimessenger observations of a flaring blazar coincident with high-energy neutrino IceCube-170922A,''
Science \textbf{361}, no.6398, eaat1378 (2018)
doi:10.1126/science.aat1378
[arXiv:1807.08816 [astro-ph.HE]].

\bibitem{IceCube:2018cha}
M.~G.~Aartsen \textit{et al.} [IceCube],
``Neutrino emission from the direction of the blazar TXS 0506+056 prior to the IceCube-170922A alert,''
Science \textbf{361}, no.6398, 147-151 (2018)
doi:10.1126/science.aat2890
[arXiv:1807.08794 [astro-ph.HE]].

\bibitem{Kelly:2018tyg}
K.~J.~Kelly and P.~A.~N.~Machado,
``Multimessenger Astronomy and New Neutrino Physics,''
JCAP \textbf{10}, 048 (2018)
doi:10.1088/1475-7516/2018/10/048
[arXiv:1808.02889 [hep-ph]].

\bibitem{Ioka:2014kca}
K.~Ioka and K.~Murase,
``IceCube PeV\textendash{}EeV neutrinos and secret interactions of neutrinos,''
PTEP \textbf{2014}, no.6, 061E01 (2014)
doi:10.1093/ptep/ptu090
[arXiv:1404.2279 [astro-ph.HE]].

\bibitem{Bustamante:2020mep}
M.~Bustamante, C.~Rosenstr\o{}m, S.~Shalgar and I.~Tamborra,
``Bounds on secret neutrino interactions from high-energy astrophysical neutrinos,''
Phys. Rev. D \textbf{101}, no.12, 123024 (2020)
doi:10.1103/PhysRevD.101.123024
[arXiv:2001.04994 [astro-ph.HE]].

\bibitem{IceCube:2022der}
R.~Abbasi \textit{et al.} [IceCube],
``Evidence for neutrino emission from the nearby active galaxy NGC 1068,''
Science \textbf{378}, no.6619, 538-543 (2022)
doi:10.1126/science.abg3395
[arXiv:2211.09972 [astro-ph.HE]].

\bibitem{IceCube:ngc1068-data}
IceCube Collaboration,
``Evidence for neutrino emission from the nearby active galaxy NGC 1068. Dataset.'' (2022)
doi:10.1126/10.21234/03fq-rh11

\bibitem{Cline:2022qld}
J.~M.~Cline, S.~Gao, F.~Guo, Z.~Lin, S.~Liu, M.~Puel, P.~Todd and T.~Xiao,
``Blazar Constraints on Neutrino-Dark Matter Scattering,''
Phys. Rev. Lett. \textbf{130}, no.9, 091402 (2023)
doi:10.1103/PhysRevLett.130.091402
[arXiv:2209.02713 [hep-ph]].

\bibitem{Ferrer:2022kei}
F.~Ferrer, G.~Herrera and A.~Ibarra,
``New constraints on the dark matter-neutrino and dark matter-photon scattering cross sections from TXS 0506+056,''
JCAP \textbf{05}, 057 (2023)
doi:10.1088/1475-7516/2023/05/057
[arXiv:2209.06339 [hep-ph]].

\bibitem{Cline:2023tkp}
J.~M.~Cline and M.~Puel,
``NGC 1068 constraints on neutrino-dark matter scattering,''
[arXiv:2301.08756 [hep-ph]].

\bibitem{Stecker:1991vm}
F.~W.~Stecker, C.~Done, M.~H.~Salamon and P.~Sommers,
``High-energy neutrinos from active galactic nuclei,''
Phys. Rev. Lett. \textbf{66}, 2697-2700 (1991)
[erratum: Phys. Rev. Lett. \textbf{69}, 2738 (1992)]
doi:10.1103/PhysRevLett.66.2697

\bibitem{Esteban:2020cvm}
I.~Esteban, M.~C.~Gonzalez-Garcia, M.~Maltoni, T.~Schwetz and A.~Zhou,
``The fate of hints: updated global analysis of three-flavor neutrino oscillations,''
JHEP \textbf{09}, 178 (2020)
doi:10.1007/JHEP09(2020)178
[arXiv:2007.14792 [hep-ph]].
NuFIT 5.2 (2022), www.nu-fit.org.

\bibitem{Planck:2018vyg}
N.~Aghanim \textit{et al.} [Planck],
``Planck 2018 results. VI. Cosmological parameters,''
Astron. Astrophys. \textbf{641}, A6 (2020)
[erratum: Astron. Astrophys. \textbf{652}, C4 (2021)]
doi:10.1051/0004-6361/201833910
[arXiv:1807.06209 [astro-ph.CO]].

\bibitem{Braun:2008bg}
J.~Braun, J.~Dumm, F.~De Palma, C.~Finley, A.~Karle and T.~Montaruli,
``Methods for point source analysis in high energy neutrino telescopes,''
Astropart. Phys. \textbf{29}, 299-305 (2008)
doi:10.1016/j.astropartphys.2008.02.007
[arXiv:0801.1604 [astro-ph]].

\bibitem{Murase:2019xqi}
K.~Murase and I.~M.~Shoemaker,
``Neutrino Echoes from Multimessenger Transient Sources,''
Phys. Rev. Lett. \textbf{123}, no.24, 241102 (2019)
doi:10.1103/PhysRevLett.123.241102
[arXiv:1903.08607 [hep-ph]].

\bibitem{Cowan:1998ji}
G.~Cowan,
``Statistical data analysis,'' Oxford University Press (1998)

\bibitem{Murase:2022feu}
K.~Murase and F.~W.~Stecker,
``High-Energy Neutrinos from Active Galactic Nuclei,''
[arXiv:2202.03381 [astro-ph.HE]].

\bibitem{Creque-Sarbinowski:2021nil}
C.~Creque-Sarbinowski, M.~Kamionkowski and B.~Zhou,
``Seeking neutrino emission from AGN through temporal and spatial cross-correlation,''
Phys. Rev. D \textbf{105}, no.12, 123035 (2022)
doi:10.1103/PhysRevD.105.123035
[arXiv:2111.08012 [astro-ph.HE]].

\bibitem{Doring:2023vmk}
C.~D\"oring and S.~Vogl,
``Astrophysical neutrino point sources as a probe of new physics,''
[arXiv:2304.08533 [hep-ph]].

\bibitem{Braun:thesis}
J.~Braun,
``A Maximum-Likelihood Search for Neutrino Point Sources with the AMANDA-II Detector,'' PhD Thesis (2009)

\bibitem{IceCube:2021xar}
R.~Abbasi \textit{et al.} [IceCube],
``IceCube Data for Neutrino Point-Source Searches Years 2008-2018,''
doi:10.21234/CPKQ-K003
[arXiv:2101.09836 [astro-ph.HE]].


\end{thebibliography}
\end{document}